\RequirePackage{fix-cm}
\documentclass[smallextended]{svjour3}       
\smartqed  
\usepackage{graphicx}
\usepackage{array}
\usepackage{wrapfig}
\usepackage{float}
\usepackage{multirow}
\usepackage{amsmath}
\usepackage{siunitx}
\usepackage{longtable,tabularx}
\usepackage{subfloat}
\usepackage{subfigure}
\usepackage{txfonts}
\usepackage{appendix}

\newlength{\tempheight}
\newlength{\tempwidth}
\newcommand{\rowname}[1]
{\rotatebox{90}{\makebox[\tempheight][c]{\textbf{#1}}}}
\newcommand{\columnname}[1]
{\makebox[\tempwidth][c]{\textbf{#1}}}

\DeclareSymbolFont{matha}{OML}{txmi}{m}{it}

\allowdisplaybreaks

\begin{document}

\title{An In-Plane $J_2$-Invariance Condition and Control Algorithm for Highly Elliptical Satellite Formations}

\author{Jackson Kulik}


\institute{J. Kulik \at
              The Aerospace Corporation \\
              \email{jackson.kulik@aero.org, jpk258@cornell.edu}           
}

\date{Received: date / Accepted: date}

\titlerunning{An In-Plane $J_2$-Invariance Condition}

\maketitle

\begin{abstract}
Phased formations of satellites provide an important means of keeping multiple satellites in close proximity without becoming dangerously close to one another. In order to minimize the amount of propellant necessary to keep a set of satellites in a phased formation, this paper presents a new condition for in-plane linearized secular $J_2$-invariance in highly eccentric orbits. A maintenance strategy outlined here combats the growth of out-of-plane motion, and the delta-v cost of this strategy is analyzed. For context, this paper also discusses relevant formations that benefit from this condition and maintenance strategy. A highly elliptical perching formation called the ``boomerang perch" is also discussed.

\keywords{Elliptical Formation Flight \and Perturbations \and Relative Motion \and J2}
\end{abstract}

\section{Introduction}
Highly elliptical orbits such as Tundra or Molniya orbits are useful for scientific and communication satellites as a result of their high altitudes or their long dwell times over apogee \cite{draim2004satellite}. Formation flight might be useful in these contexts for situational awareness, redundancy, or scientific measurement purposes. Highly elliptical formation flight has been showcased in the NASA Magnetospheric Multiscale Mission (MMS) to study the Earth's magnetosphere with a formation that has reached inter-satellite distances as low as tens of kilometers in a tetrahedral formation \cite{magneto,roscoe}. In these contexts, Earth's gravitational harmonics become a significant perturbation given the low perigee altitudes of highly elliptical orbits. As such, formations that can reduce the influences of these perturbations while maintaining design flexibility are desirable.

We shall be intentionally vague in our definition of what constitutes slightly versus highly elliptical. However, in the context of this paper, one can be sure that an orbit with $e<0.01$ is slightly eccentric, while an orbit with $e>0.3$ is considered highly elliptical.  The author acknowledges an ambiguous region between eccentricity values of 0.01 and 0.3, which shall be referred to in this paper as moderately elliptical.

Current methods for designing formations with perturbations in mind do not generalize well for highly elliptical orbits. Schaub and Alfriend \cite{schaub2001J2} identified a class of formations for which the $J_2$ perturbation causes equal but opposite growth in certain mean classical element differences. This paper will show that while this class of formations requires less propellant to maintain than other formations with similar geometric properties in slightly elliptical contexts, the opposite is true in highly elliptical contexts.

The analytical methods of Schaub and Alfriend have been extended by many researchers to explore alternative constraints on the linearized secular growth of orbital elements as a result of $J_2$ \cite{alfriend2009spacecraft,breger2006partial,dang2015}, or on the mean inter-satellite distance \cite{nie2018analytical}. Other numerical and dynamical systems approaches to formation establishment such as \cite{baresi2017bounded,baresi2017design,he2018bounded,lara2012integrable,martinucsi2011solutions,xu2012existence} avoid the use of mean elements and generally account for arbitrary zonal gravitational harmonics. However, according to Baressi and Scheeres, these dynamical systems techniques do not readily extend to highly elliptical formation design due to the non-existence of pseudo-circular orbits matching the nodal and sidereal periods of orbits with certain combinations of eccentricity and inclination \cite{baresi2017bounded,baresi2017design,broucke1994numerical}.

In order to reap the benefits of formation flight in highly elliptical regimes, it is necessary to identify classes of formations in this domain with diverse geometric properties but little propellant cost to maintain. This paper examines a class of formations that practically maintain their periodic in-plane motion despite the perturbation from linearized secular $J_2$. We develop the constraints determining membership within this class of formations from a differential mean orbital element perspective. The condition derived here is distinct from any of the other analytical partial $J_2$-invariant conditions discussed so far. Within this class of formations, numerical results are presented to study and adjust formation initial conditions to additionally reduce some of the influence of arbitrary gravitational harmonics. We make modifications to the formation-keeping strategy outlined by Schaub and Alfriend \cite{impulse} to particularly suit this class of formations, and the type of drift they tend to experience. Numerical results are presented to demonstrate the efficacy of this class of formations with the accompanying maintenance strategy. The ``boomerang perch" formation is also introduced, and used as an example of delta-v savings achieved by the formation maintenance algorithm derived here on non-wheel type formations.

The emphasis in this paper is on constructing wheel-type formations that consist of a single chief satellite surrounded by multiple deputy satellites in approximately the same sized natural motion circumnavigation (NMC)/``wheel," but offset from one another by typically equal phase angles. These formations allow for many satellites to maintain fairly consistent separation from one another while circumnavigating the chief satellite. ``Wheel" is used rather than NMC since it implies the importance of placing multiple deputies at different phase angles. Additionally, ``NMC" brings with it the connotation of the 2x1 ellipse, which is not consistent with the natural relative motion of formations with elliptical reference orbits since these possess asymmetries and deviate from the 2x1 ratio.  Along the way to describing and simulating the various classes of formations described above, we elaborate upon existing methods for geometric formation design of wheel formations. This serves to derive, simplify, and correct a minor error in the formation design work of Chao \cite{chao2017deployment,chao2005applied}, and to summarize work by Sengupta and Vadali \cite{sengupta2007}. This paper's correction and expansion of the work of Chao gives a generalized inversion of the low reference eccentricity equations of linearized relative motion presented in \cite{alfriend2009spacecraft}. The formation design equations from \cite{alfriend2009spacecraft} itself focus on establishing projected circular orbit (PCO) and general circular orbit (GCO) formations, whereas this paper provides novel expressions for the design of the in-plane component of other general formation geometries. Between the focus on geometric design and perturbation invariance, this paper serves as a starting point for the intuitive design of robust wheel formations in highly elliptical regimes. The original low eccentricity perturbation invariance conditions and formation design methods are presented and then followed by their highly elliptical counterparts for comparison.

\section{Introduction to $J_2$ Invariance}
 Using Lagrange's variation of parameters technique \cite{vallado2001fundamentals}, it is possible to derive the linearized approximations for the secular rates of change for all six mean orbital elements ($a,e,i,\Omega,\omega,M$; semi-major axis, eccentricity, inclination, right ascension of the ascending node, argument of perigee, and mean anomaly respectively) due to the $J_2$ perturbation. Though analysis will be conducted in radians, orbital elements and differential orbital elements will be given in degrees and kilometers. Three of the mean orbital elements have a nonzero secular first-order rate of change as a result of the $J_2$ term in the gravitational potential of the Earth:
\begin{equation}
\label{draan}
\dot{\Omega}=-\frac{3nR_e^2J_2}{2p^2}\cos i
\end{equation}
\begin{equation}
\label{dw}
\dot{\omega}=\frac{3nR_e^2J_2}{4p^2}(4-5\sin^2i)
\end{equation}
\begin{equation}
\label{dM}
\dot{M}=n-\frac{3nR_e^2J_2\sqrt{1-e^2}}{4p^2}(3\sin^2i-2)
\end{equation}
where $n$ is the mean motion, $p=a(1-e^2)$ is the parameter of the ellipse, $R_e$ is the radius of the Earth, and $J_2$ is a coefficient in the series expansion for Earth's gravitational potential.

The strictest form of $J_2$-invariance is a formation in which all satellites share all osculating elements besides right ascension of the ascending node. In this very narrow case, all satellites are at the same latitude at all times, and the acceleration due to $J_2$ is the same for each satellite. By this symmetry, the osculating differential orbital elements remain constant in a model with two body and $J_2$ interactions. This is quite restrictive, so one looks instead to mean elements and the effects of linearized secular $J_2$. 

In general, to establish a formation of multiple satellites that exhibit periodic relative motion despite the linearized secular approximation to the $J_2$ perturbation, one must either set all variables on the right hand side of equations \ref{draan}-\ref{dM} equal between each satellite in the formation, or make some compromises on exactly what qualities of the formation are invariant to $J_2$. The first choice implies

\begin{equation}
\label{zeros}
\delta a=\delta e=\delta i=0
\end{equation}
which still allows three degrees of freedom when designing relative orbits (and even allows for the design of certain wheel type formations  in highly elliptical reference orbits). However, it is often desirable in mission design to make use of other differential orbital elements. As such, compromises may be made to open up one or more degrees of freedom in the formation design space. For example, it is desirable in the construction of phased wheel formations of multiple deputy satellites \cite{chao2017deployment,chao2005applied} to be able to use $\delta e$ as a design parameter.

As a result, Schaub and Alfriend \cite{alfriend2009spacecraft,schaub2001J2} introduced approximate conditions for a type of $J_2$-invariance that performs well (in the sense of limiting the drift between chief and deputy) for circular and slightly elliptical orbits and allows for one of the three differential elements from equation \ref{zeros} to be chosen as a non-zero quantity.

The principle upon which their work is based is the following set of conditions:

\begin{align}
\dot{\Omega}_d-\dot{\Omega}_c=\delta\dot{\Omega}&=0\\
\label{alf_const}
\dot{\theta}_d-\dot{\theta}_c=\dot{\omega}_d+\dot{M}_d-(\dot{\omega}_c+\dot{M}_c)=\delta\dot{\theta}&=0
\end{align}
that the ascending nodes of the two satellites do not drift apart, and that the mean arguments of latitude ($\theta=M+\omega$) also remain the same as one another. For circular or slightly elliptical orbits, relative motion remains fairly unchanged by $J_2$ when these conditions are satisfied. 

In \cite{alfriend2009spacecraft}, the second condition above (equation \ref{alf_const}) was replaced by a more generic ``no in-track drifting condition" that incorporates the in-track effects of $\cos (i)\delta\dot{\Omega}$, and a third condition was presented in which $\delta\dot{\omega}$ was set to zero. The authors largely dismissed the third condition as useless, and it should be noted that the no differential RAAN growth condition when combined with the ``no in-track drift condition" yields the same equations as above from the original paper \cite{schaub2001J2}. The final constraints on $\delta a$ and $\delta i$ that result from the analyses are identical despite being presented in nondimensional coordinates in \cite{schaub2001J2} and dimensional coordinates in \cite{alfriend2009spacecraft}. It should also be noted that the ``no in-track drifting condition" is something of a misnomer in highly elliptical reference orbits. Note the asymmetry in the in-track bounds of figure \ref{fig:eq} that develops over time despite the adherence to the ``no in-track drifting condition." This is addressed in \cite{dang2015} by the addition of a correction factor $\beta$ in the relation between the growth of $\delta M$ and $\delta \omega$, ensuring the symmetry of the in-track bounds despite the formation's growth in size. Despite maintaining the centering of the formation, this approach still fails to prevent expansion in size of highly elliptical wheels.

\section{Counterexample for Mean Argument of Latitude Constraint} \label{cx}
Here, a domain in which the conditions of equation \ref{alf_const} do not produce satisfactory periodic relative motion is presented. The condition presented in equation \ref{alf_const} implies that
\begin{equation}
\delta\dot{\omega}=-\delta\dot{M}
\end{equation}

\begin{figure}[hbt!] 
	\centering
	\subfigure {\includegraphics[width=0.45\textwidth]{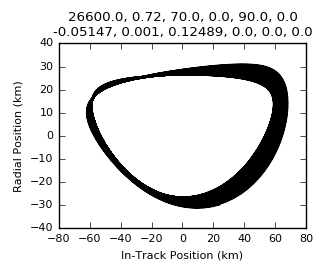}} 
	\subfigure {\includegraphics[width=0.45\textwidth]{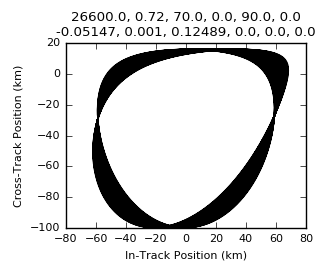}}\\
	\caption{These orbits are calculated with the procedure outlined in \cite{schaub2001J2}. Fifty periods of the chief orbit are pictured. Propagation includes first-order secular $J_2$ effects. The titles represent the mean orbital elements $(a,e,i,\Omega,\omega,M)$ of the chief in the first line, and the differential mean elements of the deputy on the next line. All dimensional terms are in kilometers and degrees.}
	\label{fig:eq}
\end{figure}

This implies that the condition allows for the arguments of perigee of two orbits to drift apart as long as their relative mean anomalies compensate in an equal and opposite manner. In a situation where the two orbits are highly eccentric, this will cause a large change in the relative motion over time. To bring this to an extreme, imagine two coplanar highly elliptical orbits that began with the same argument of perigee and mean anomaly. Over time, they have been affected by $J_2$ for so long that the perigees have rotated $180$ degrees from one another, and the mean anomalies have adjusted in the opposite manner. When one satellite is at perigee, the other is at apogee, and the two satellites are $2ae$ apart from one another, on the order of the semi-major axis for a high eccentricity. On the other hand, for slightly elliptical orbits, opposing perigees with correspondingly modified mean anomalies can result in relative motion confined to a small region, so that the behavior stemming from equation \ref{alf_const} is more reasonable. This is why slightly elliptical $J_2$-invariant orbits do not grow much in size, but the formation depicted in figure \ref{fig:eq} does.

Pictured in figure \ref{fig:eq} is the relative motion of two satellites that approximately meet the constraints from equation \ref{alf_const}. The simulation is propagated with linearized secular $J_2$ effects in mean element space along the lines of the pkepler algorithm from \cite{vallado2001fundamentals}. This low fidelity simulation is justifiable given that it relies on the same linearized secular expressions from which the various $J_2$-invariance constraints were developed. Essentially, these low fidelity simulations show exactly how the relative orbits are designed to act and approximately how they will act in a higher fidelity simulation. Visibly, the relative motion in figure \ref{fig:eq} is not qualitatively $J_2$-invariant. While this method works quite well for orbits with lower eccentricities, it is clear that a new method must be developed if one wishes to use $\delta e$ in the design of highly elliptical formations that are meant to be $J_2$-invariant in the sense of non-drifting relative motion.

\section{Development of an Alternative Constraint}
Here, a condition is presented to allow for an in-plane $J_2$-invariant relative motion even in highly elliptical contexts. This condition is motivated by the observation that for small $\delta\Omega$ and $\delta\omega$, the linearized forms of the relative motion have in-plane terms of a similar form \cite{sengupta2007,schaub2004}.
\begin{figure}[hbt!] 
	\centering
    \includegraphics[width=0.7\textwidth]{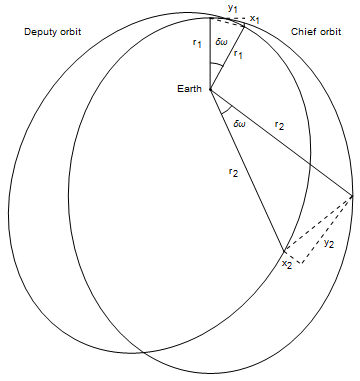}
	\caption{Two elliptical orbits about the Earth sharing all orbital elements besides argument of perigee. The in-track and radial displacements are shown at two different times.}
	\label{fig:rect_draw}
\end{figure}
\begin{figure}[hbt!] 
	\centering
	\subfigure {\includegraphics[width=0.45\textwidth]{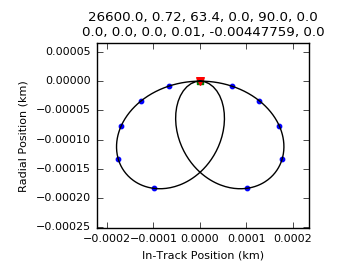}} 
	\subfigure {\includegraphics[width=0.45\textwidth]{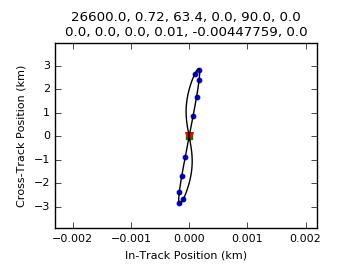}}\\
	\caption{A two-body propagated relative orbit almost confined to the cross-track axis. Twelve equally time-spaced dots are pictured. The deputy is at the origin at the two apses.}
	\label{fig:rect}

\end{figure}

For relative motion induced by small $\delta\omega$:
\begin{align}
\label{rectilinear-parametric}
x&=r(\cos\delta\omega-1)\approx0\\
y&=r\sin\delta\omega\approx r\delta\omega\\
z&=0
\end{align}
where $x,y,z$ are the radial, in-track, and cross-track distances between chief and deputy, and $\nu$ is the chief true anomaly. The exact equalities in the above equation go beyond the given references' more general but approximate relations regardless of the size of $\delta\omega$. The exactness of this special case can easily be proved from a geometric argument given the information that the angle between the two satellites is always $\delta\omega$, and their distance from the center of the Earth is the same at all times. Pictured in figure \ref{fig:rect_draw} are the similar isosceles triangles that are formed by the Earth and the two satellites. Thus, the relative motion is a perfect line that falls off from the in-track axis by $\delta\omega/2$.  For relative motion induced by small $\delta\Omega$:

\begin{align}
\label{raan-parametric}
x&\approx 0\\
y&\approx r\cos(i)\delta\Omega\\
z&\approx -r\cos(\nu+\omega)\sin(i)\delta\Omega
\end{align}

As such, the following condition leads to an almost exclusively cross-track relative motion that is pictured in figure \ref{fig:rect}.
\begin{equation}
\label{cross_rel}
\delta\omega= -\cos(i)\delta\Omega
\end{equation}

With this in mind, a similar constraint is placed on the growth of the orbital elements to produce an in-plane $J_2$-invariant relative motion:

\begin{align}
\label{Mconst}
\delta\dot{M}=\dot{M}_d-\dot{M}_c=0\\
\label{crossconst}
\delta\dot{\omega}=\dot{\omega}_d-\dot{\omega}_c=-\cos(i_c)\Big(\dot{\Omega}_d-\dot{\Omega}_c\Big)=-\cos(i_c)\delta\dot{\Omega}
\end{align}

Taking the first order Taylor series in $\delta a, \delta e, \delta i$, equations \ref{Mconst} and \ref{crossconst} simplify to equations \ref{Mconst1} and \ref{crossconst1}, respectively, where $\eta=\sqrt{1-e^2}$

\begin{align}
\label{Mconst1}
6J_2R^2ae(2-3\sin^2i)\delta e-[4a^2\eta^{5}+7J_2R^2\eta^2(2-3\sin^2i)]\delta a-6J_2R^2a\eta^2\sin(2i)\delta i&=0\\
\label{crossconst1}
8ae(2-3\sin^2i)\delta e-7\eta^2(2-3\sin^2i)\delta a-8a\eta^2\sin(2i)\delta i&=0
\end{align}

The solution for the linear system given by equations \ref{Mconst1} and \ref{crossconst1} is

\begin{align}
\label{da1}
\delta a&=0\\
\delta i&=\frac{(2-3\sin^2i)e\delta e}{(1-e^2)\sin(2i)}
\label{di1}
\end{align}
Alternatively, one can solve equations \ref{Mconst} and \ref{crossconst} with a numerical root finding scheme. The root finding problem can be simplified to one equation and one unknown by solving for $\sin^2i_d$ in equation \ref{Mconst} (which references equation \ref{dM}), from which $\cos^2i_d$ can be deduced. Assuming the sign of $\cos i_d$ matches the sign of $\cos i_c$, and substituting these expressions into equation \ref{crossconst} gives an equation for which the only unknown is $\delta a$. After solving numerically for $\delta a$, this value can be substituted back into the expression for $\cos i_d$, yielding the value of $\delta i$. The exact expressions are cumbersome and omitted.

\begin{figure}[hbt!] 
	\centering
	\subfigure {\includegraphics[width=0.45\textwidth]{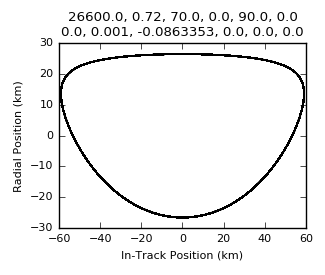}} 
	\subfigure {\includegraphics[width=0.45\textwidth]{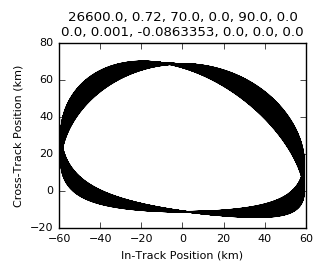}}\\
	\caption{One hundred periods of a relative motion satisfying equation \ref{di1}. Propagation with first-order secular J2 effects.}
	\label{fig:inv}
\end{figure}

An example of a relative orbit (with a highly eccentric $e=0.72$ chief) calculated by using this constraint is shown in figure \ref{fig:inv}. The differential orbital elements were chosen by selecting $\delta e=0.001$ as a design parameter. From this choice, $\delta i$ was determined by using equation \ref{di1}, and $\delta a$ was set to zero according to equation \ref{da1}. The three remaining differential orbital elements were also free to be chosen at will, but were set to $0$ for simplicity. The in-plane motion is very nearly unchanged despite perturbations, while the cross-track motion changed significantly in its phase and amplitude over the course of the simulation.

\section{Comparison of Eccentricity Induced Relative Orbits}

In figure 5, three different methods are shown for producing eccentricity induced relative motions: $\delta e$ only, mean latitude method (of Schaub and Alfriend \cite{schaub2001J2}), and the in-plane invariant method (derived in this paper).  Simulations are run in two contexts, a moderate eccentricity ($e=0.1$) LEO orbit, and a high eccentricity ($e=0.72$) Molniya orbit. Note that the in-plane dimensions of any of the three formations being compared start the same since they share $\delta e$, making this the uniform standard across which the three methods are compared. Further, the differences in cross-track motion amplitude between formations is due to the constrained differential inclination, which is a function of the differential eccentricity. Differential right ascension of the ascending node could be employed to make all cross-track amplitudes uniform across the comparison. However, since propagation is performed with linearized secular $J_2$, adjustments to $\delta\Omega$ have no bearing on the amount of drift in the relative motion, and are not included for simplicity of the comparison. Therefore, differences in the size of the formations being compared either result from perturbations during the course of the simulation, or do not have an impact on the amount of drift each formation experiences.
\begin{figure}[hbt!]
\label{comp}
	\setlength{\tempwidth}{.32\linewidth}
	\settoheight{\tempheight}{\includegraphics[width=\tempwidth]{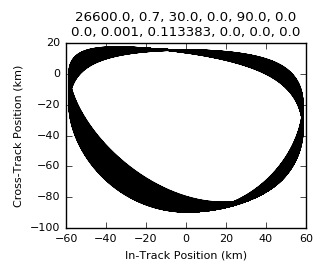}}
	\centering
	\hspace{\baselineskip}
	\columnname{$\delta e$ Only}\hfil
	\columnname{Mean Latitude Method}\hfil
	\columnname{In-Plane Invariant Method}\\
	\rowname{In-Plane LEO}
	\subfigure{\includegraphics[width=\tempwidth]{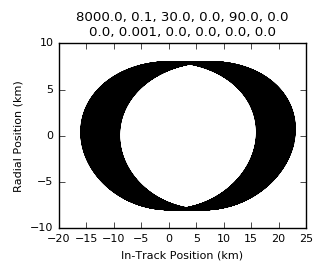}}\hfil
	\subfigure{\includegraphics[width=\tempwidth]{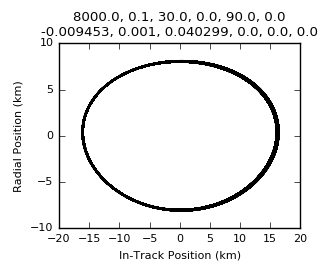}}\hfil
	\subfigure{\includegraphics[width=.33\linewidth]{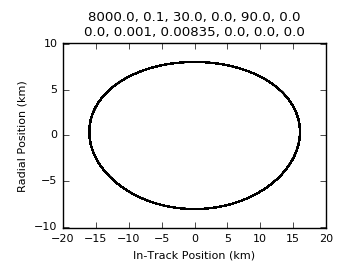}}\\
	\rowname{Out of Plane LEO}
	\subfigure{\includegraphics[width=\tempwidth]{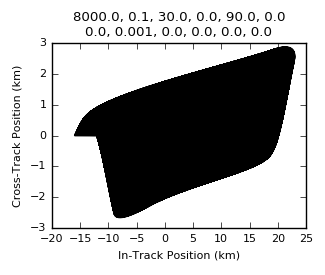}}\hfil
	\subfigure{\includegraphics[width=\tempwidth]{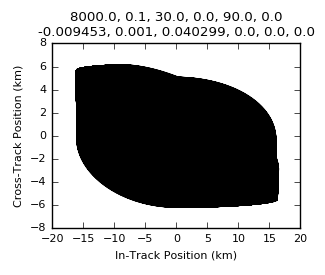}}\hfil
	\subfigure{\includegraphics[width=\tempwidth]{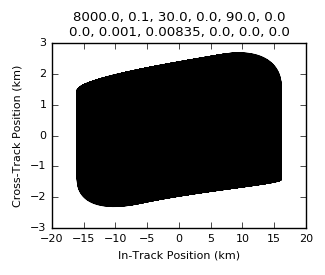}}\\
	\rowname{In-Plane Molniya}
	\subfigure{\includegraphics[width=\tempwidth]{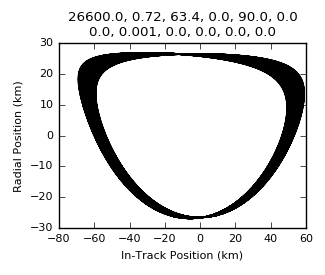}}\hfil
	\subfigure{\includegraphics[width=\tempwidth]{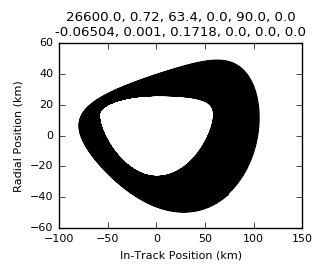}}\hfil
	\subfigure{\includegraphics[width=\tempwidth]{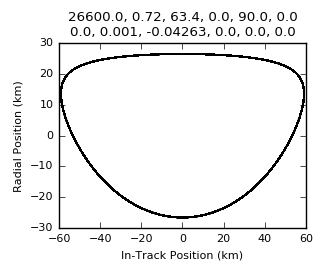}}\\
	\rowname{Out of Plane Molniya}
	\subfigure{\includegraphics[width=\tempwidth]{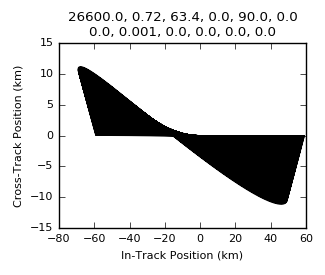}}\hfil
	\subfigure{\includegraphics[width=\tempwidth]{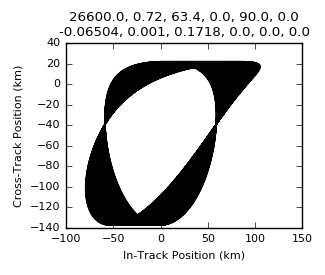}}\hfil
	\subfigure{\includegraphics[width=\tempwidth]{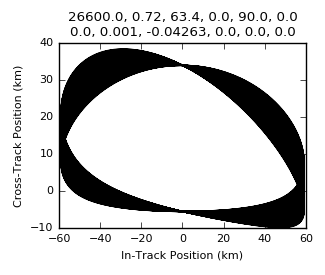}}
	\caption{Plots of the in-plane and out of plane components of six different relative orbits under linearized secular $J_2$ propagation. LEO cases propagated for 300 periods, and Molniya cases propagated for 100 periods.}
	\label{fig:fig3}
\end{figure}

It can be seen that in both of these contexts, the $J_2$ effects cause a significant and undesirable drift in the relative motion for the simple $\delta e$ only case, or for the mean latitude condition based relative orbits. This is because the $J_2$ effect is most pronounced for high eccentricity or for low semi-major axis (see equations \ref{draan},\ref{dw},\ref{dM}). The in-plane drift is highest in the high eccentricity mean latitude based relative orbit, because the condition causes a drift in perigees that leads to the most pronounced change in relative motion for a high eccentricity reference orbit. 

Note that the mean latitude condition cross-track extrema do unexpectedly change in size despite $\delta\Omega$ and $\delta i$ being constant throughout the simulation. These changes in cross-track motion are induced by the drift of the reference orbit argument of perigee that affects the phase of the motion induced by the constant differential inclination. This drift in reference argument of perigee does not affect the normalized amplitude ($Q_3$) of the cross-track motion from \cite{sengupta2007} (cross-track motion divided by the reference satellite's distance from the Earth). However, the reference perigee drift does affect the actual extrema of the cross-track motion, since the two oscillating quantities' (chief distance from Earth, and the harmonic oscillator in the normalized coordinates) phases are changed relative to one another.

\section{Application to Phased Wheel Formations}

While potentially useful on its own, the in-plane $J_2$-invariant condition was originally motivated by the problem of establishing multiple satellites in a wheel-like formation such as the one developed by Chao in \cite{chao2017deployment,chao2005applied}. To initialize a formation of satellites in roughly the same size wheel, but at different phase angles around one chief satellite, one can follow Chao's algorithm (assuming that the chief satellite is at perigee):

\begin{align}
\Delta e &= D/a\\
e_i&=(e^2+\Delta e^2-2e\Delta e\cos\beta_i)^{1/2}\\
S_i&=\frac{\Delta e}{e_i}\sin\beta_i\\
\omega_i&=\sin^{-1}(S_i)+\omega\\
M_i&=2\pi-\sin^{-1}(S_i)
\end{align}

where $D$ is the semi-minor axis of the formation, and $\beta_i$ is the angle counter-clockwise from the radial upwards axis for the $i$th satellite. $S_i$ is simply an intermediate variable employed by Chao.

Note that this algorithm must be modified for certain phase angles and choices of wheel size for very slightly eccentric orbits. For example, choose $D$ large enough such that $\Delta e$ is greater than $e$ of chief, and choose $\beta=0$, then the algorithm will break down and the wheel size will not be $D$ as intended. This behavior arises from the fact that $\sin^{-1}(S_i)\in[-\pi/2,\pi/2]$ radians, when at times the only way to accomplish a given wheel at a certain phase angle is to have the deputy argument of perigee be more than $\pi/2$ radians off from the chief perigee.

In order to remedy this, it is necessary to modify the algorithm such that $\omega_i$ can take the value of any angle. One can study $S_i$ around the value $1$. $S_i=1$ implies that
\begin{equation}
\Delta e=\pm\frac{e}{\cos\beta_i}
\end{equation}
One can ignore the negative solution since $\Delta e$ is assumed to be positive in this algorithm. As such, when $\Delta e>\frac{e}{\cos\beta_i}$, one can assume that $\sin^{-1}(S_i)$ should actually be in the interval $(\pi/2,3\pi/2)$ radians.

As such, the modified algorithm becomes as follows

\begin{align}
\Delta e &= D/a\\
e_i&=(e^2+\Delta e^2-2e\Delta e\cos\beta_i)^{1/2}\\
S_i&=\frac{\Delta e}{e_i}\sin\beta_i\\
\omega_i &=
\begin{cases}
\sin^{-1}(S_i)+\omega        & \text{if } \cos\beta_i\leq0 \text{ or } \Delta e\leq\frac{e}{\cos\beta_i} \\
\pi-\sin^{-1}(S_i)+\omega        & \text{otherwise}
\end{cases}\\
M_i &=
\begin{cases}
2\pi-\sin^{-1}(S_i)        & \text{if } \cos\beta_i\leq0 \text{ or } \Delta e\leq\frac{e}{\cos\beta_i} \\
\pi+\sin^{-1}(S_i)       & \text{otherwise}
\end{cases}
\end{align}

Or equivalently, derived in the appendix with arctan2(y,x):

\begin{align}
\Delta e &= D/a\\
e_i&=(e^2+\Delta e^2-2e\Delta e\cos\beta_i)^{1/2}\\
\omega_i &= \omega-\mathrm{arctan2}\big(-D\sin\beta_i,ae_c-D\cos\beta_i\big) \\
M_i &= \mathrm{arctan2}\big(-D\sin\beta_i,ae_c-D\cos\beta_i\big)
\end{align}

Following the low eccentricity linearized equations of relative motion from \cite{alfriend2009spacecraft} page 213, one might also consider correcting small spatial biases by increasing the deputy argument of perigee by $e(e_i\sin M_i)$, but this is not pursued further given the small size of this correction for small eccentricities.

Suppose that a chief satellite has element set (9000km, 0.002, 0, 0, 0, 0) and that a deputy satellite is being placed in a formation with $D=40$, $\beta=45$ degrees. The original algorithm will place the deputy at the differential elements (0km, 0.001344, 0, 0, 70.018536, 289.981463), and the modified algorithm will place the deputy at the differential elements (0km, 0.001344, 0, 0, 109.981463, 250.018536) with all angular units in degrees. These deputies are plotted in figure 6.

\begin{figure}[hbt!]
	\label{fig:bad_chao} 
	\centering
	\subfigure {\includegraphics[width=0.4\textwidth]{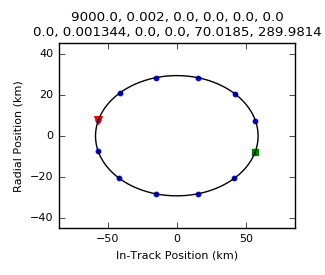}} 
	\subfigure {\includegraphics[width=0.4\textwidth]{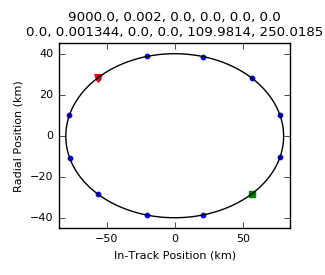}}\\
	\caption{On the left, a wheel formation created with the original Chao algorithm. Note the size and phasing of the formation are not as intended (40km instead of 30km semi-minor axis, for example). On the right, a wheel formation created with the modified Chao algorithm. Note the size and phasing of the formation are as intended. The red dot indicates the deputy location at chief perigee passage. Terms in the title reflect the relevant chief mean elements and deputy differential mean elements in kilometers and degrees.}
\end{figure}

The modified form of Chao's original algorithm for building satellite formations in the form of a phased wheel does not take into account any perturbations. It can be further improved by applying $J_2$-invariance conditions to the unconstrained differential orbital elements of the deputy satellites. For instance, in low eccentricity contexts, one might apply Schaub and Alfriend's \cite{schaub2001J2} mean latitude growth matching constraint to the deputies to modify $\delta a$ and $\delta i$ of each deputy in such a way that the relative motion will remain largely unaffected by $J_2$. As such, Chao's algorithm prescribes $\delta e,\delta\omega$, and $\delta M$, while the $J_2$-invariance condition determines $\delta a$ and $\delta i$, leaving $\delta\Omega$ as an additional free design parameter.

\section{Alternative Phased Wheel Algorithm for Highly Elliptical Contexts}

The previous section details the standard approach for constructing a $J_2$-invariant wheel in a circular to slightly elliptical regime, since the Chao algorithm and the mean latitude $J_2$-invariance condition are best suited for this regime. This paper seeks to emulate this approach but in highly elliptical regimes. The in-plane $J_2$-invariance condition suitable for highly elliptical contexts was already detailed and presented in equations \ref{da1} and \ref{di1}. However, the original formation design upon which to apply this condition has yet to be discussed. Just as the original mean latitude $J_2$-invariance condition performs poorly in highly elliptical regimes, so too does the Chao wheel builder algorithm, placing deputy satellites in formations with unintended offsets and sizes.

Using the work of Sengupta and Vadali \cite{sengupta2007}, a spatially centered highly elliptical wheel algorithm may be obtained. Given an angle $\alpha$ describing the approximate angle of the deputy from the in-track positive axis at chief perigee passage, and a wheel semi-minor axis $D$:

\begin{align}
\label{highwheel}
\delta e&=-D/a\sin\alpha\\
\delta M&=\frac{D\sqrt{1-e^2}}{ae}\cos\alpha\\
\label{highwheel1}
\delta\omega&=-\frac{D}{ae}\cos\alpha
\end{align}

Equations \ref{highwheel}-\ref{highwheel1} are derived in section \ref{A:highwheel}, and an alternative temporally centered wheel is derived and presented in section \ref{A:timewheel}. The bulk of these derivations are present in \cite{sengupta2007}, but a minor final step is given here to obtain $\delta\omega$.

While the Chao algorithm was never intended for use in highly elliptical orbits, we present a comparison that solidifies the fact that in these contexts, the alternative algorithm is necessary. Consider a chief satellite initialized in a Molniya orbit with semi-major axis 26,561km and 0.72 eccentricity. The size of the formation is specified as $D=40$km and the phase angle clockwise from radial up is $\beta=\alpha+\pi/2=3\pi/8$ radians. Then the Chao algorithm and the alternative higher eccentricity wheel algorithm are used to initialize the deputy. As shown in figure 7, the differential elements are matched almost exactly except for the differential mean anomaly. Both algorithms have approximately the same differential eccentricity and argument of perigee under the assumption of large eccentricity and small formation size, regardless of phase angle. The alternative wheel builder places the deputy in the correctly sized and centered formation, while the Chao algorithm does not.

\begin{figure}[hbt!]
    \label{chaoseng}
	\centering
	\subfigure{\includegraphics[width=.45\textwidth]{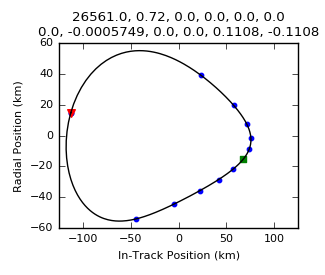}}
    \subfigure{\includegraphics[width=.45\textwidth]{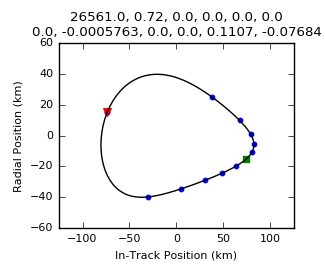}}\\
	\caption{On the left, a relative orbit designed by the Chao algorithm in an inappropriately high eccentricity regime. This formation is larger than intended, and off-center in-track. On the right, a relative orbit designed by the alternative high eccentricity algorithm that places the deputy in the desired geometry.}
\end{figure}

To create a $J_2$-invariant formation in a highly elliptical context, construct a formation using equations \ref{highwheel}-\ref{highwheel1}, and then apply equations \ref{da1} and \ref{di1}. The resulting formation will achieve the desired geometric characteristics as well as in-plane $J_2$-invariance.

According to equation \ref{highwheel}, if the phase angle is chosen such that deputy is on the in-track axis when chief is at perigee ($\alpha=0,\pi$), the deputy has no differential eccentricity with respect to the chief, and is $J_2$-invariant in the sense of equation \ref{zeros} where $\delta i = 0=\delta a$.

\section{High Fidelity Modeling}

In some Molniya and other highly elliptical orbits, simulation has shown that higher-order gravitational harmonics can exert a significant influence on the relative motion. This behavior can be explained in part by the very low perigee altitude attained and by the extremely fast variations in perturbing acceleration from these harmonics as a function of position. See for example, the slopes on an orbital perturbation chart (e.g. Figure 3.1 from page 55 of Montenbruck and Gill \cite{montenbruck2012satellite}) as evidence of the fast spatial variation of high-order gravitational harmonic perturbing accelerations. The effect is particularly strong in resonant orbits such as the Molniya with its half day period.\\

Formations established using $\delta M$ and $\delta\omega$ specifically experience a large drift over time in their differential mean anomaly from these high-order gravitational terms. Luckily, simulations demonstrate these effects can be reduced almost entirely by choosing a differential semi-major axis that will create a corresponding magnitude but opposite direction drift in differential mean anomaly. In practice, high fidelity numerical simulations have shown this approach highly effective, with differential mean semi-major axes chosen often on the order of tens or hundreds of meters (enough to correct for many kilometers of drift over the period of weeks). This remedy is similar to the differential semi-major axis correction discussed in \cite{roscoe} to account for along-track drift from lunar perturbations in the extremely high eccentricity context of phase 2 of the MMS mission.

Each of the figures presented up to this point have been produced using Keplerian or linearized secular $J_2$ propagation. To validate this work to a greater extent, we present a simulation with numerical $J_2$ as well as another two with a high fidelity 12x12 gravity model with solar and lunar gravity. In each of these three simulations, we propagate two different high eccentricity formations for a period of thirty days --- one constructed using the high eccentricity wheel algorithm (equations \ref{highwheel}-\ref{highwheel1}), and another modified according to equation \ref{di1} and then numerically adjusted for a differential semi-major axis to correct for the differential mean anomaly drift from the full numerical force model. Each formation consists of four satellites separated by $\pi/2$ radian phase angles, each with semi-minor axis of $20$km, and cross-track motion with $30$km maximum position. Cross-track motion is produced using a combination of the constrained $\delta i$ ($\delta i=0$ is chosen for the unmodified wheel) and an additional $\delta\Omega$ to attain the same maximum cross-track motion across both formations. The wheel formations in figures 8 and 9 are presented for a reference orbit with mean elements ($26561$km, $0.72$, $63.4\si{\degree}$, $70\si{\degree}$, $270\si{\degree}$, $0\si{\degree}$), while figure 10 presents wheel formations for a reference orbit with mean elements ($26000$km, $0.72$, $63.4\si{\degree}$, $70\si{\degree}$, $270\si{\degree}$, $0\si{\degree}$). Brouwer-Lyddane theory is used for converting between the mean and osculating elements prior to numerical propagation \cite{lyddane1963small}. 
The cross-track motion is of a different phase in the modified formations, since a large part of it stems from $\delta i$ rather than $\delta\Omega$. The in-plane drift is clearly reduced in the modified formations, while the out-of-plane drift is largely the same between the two formations. The in-plane drift is almost completely absent in the case of the first reference orbit propagated with numerical $J_2$, and the second reference orbit propagated with the higher fidelity model. Though less pronounced, the in-plane drift is not completely mitigated in the modified formation about the half day period resonant orbit propagated in the high fidelity model. These simulations demonstrate the utility of the in-plane $J_2$-invariance condition in resonant and non-resonant orbits to varying degrees.

\begin{figure}[hbt!]
	\label{pertmolin}
	\centering
	\subfigure{\includegraphics[width=.45\textwidth]{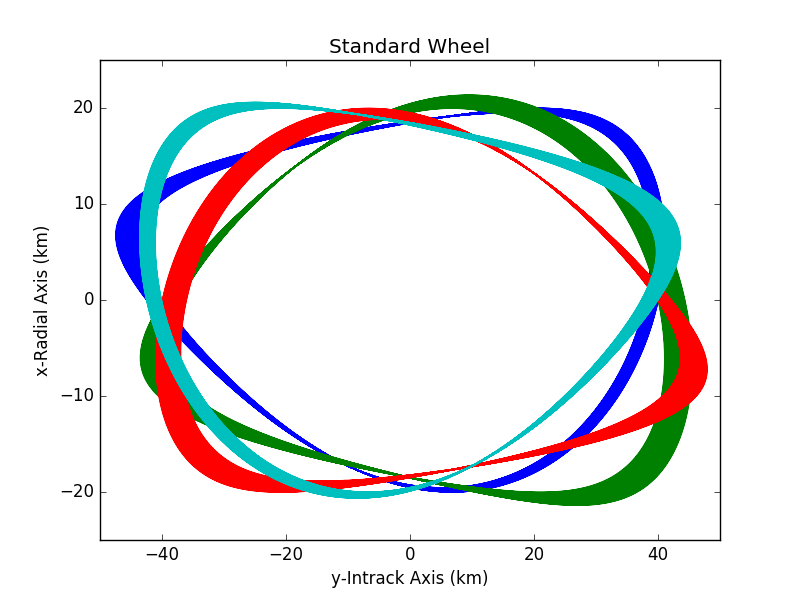}}
	\subfigure{\includegraphics[width=.45\textwidth]{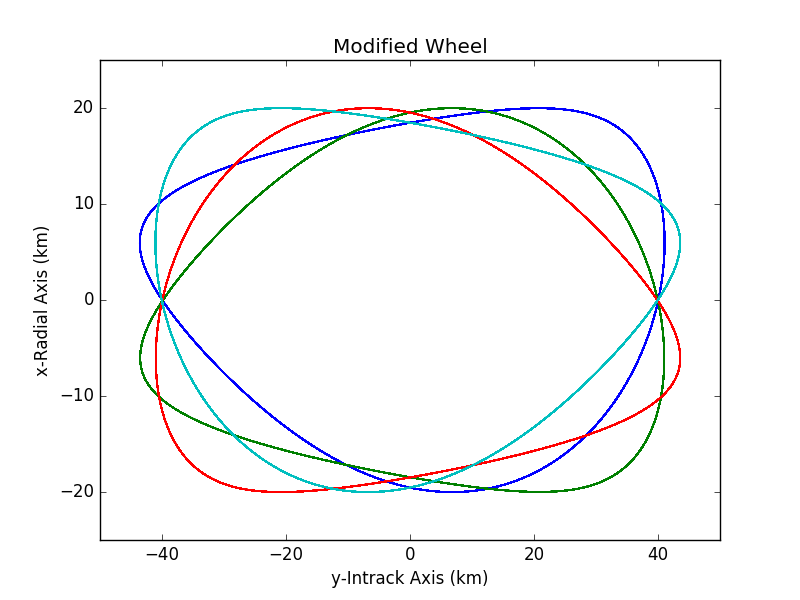}}\\
	\subfigure{\includegraphics[width=.45\textwidth]{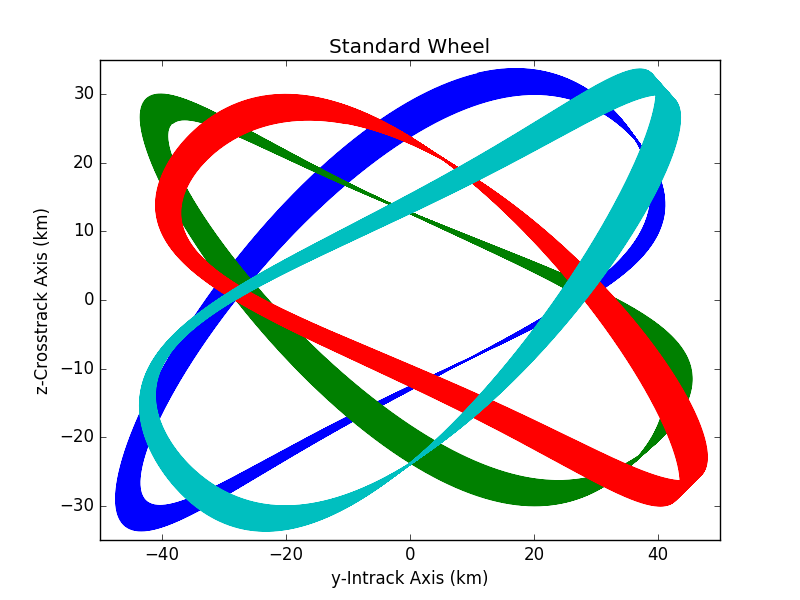}}
	\subfigure{\includegraphics[width=.45\textwidth]{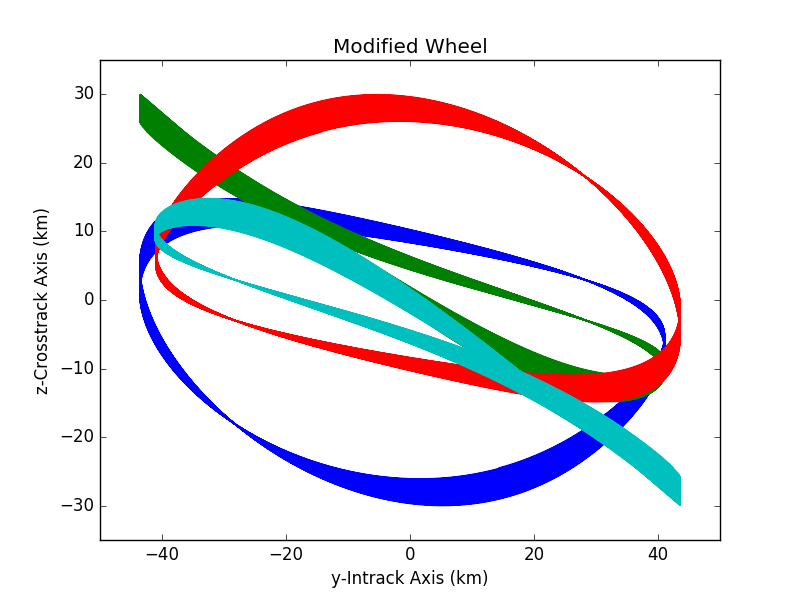}}\\
	\caption{On the left is a standard unmodified wheel formation. On the right is an in-plane $J_2$-invariant formation with numerically calculated changes to $\delta a$ to compensate for differential mean anomaly drift. Above is the in-track radial plane, and below is the in-track cross-track plane. Propagation for thirty days with numerical $J_2$.}
\end{figure}

\begin{figure}[hbt!]
	\label{pertmolin1}
	\centering
	\subfigure{\includegraphics[width=.45\textwidth]{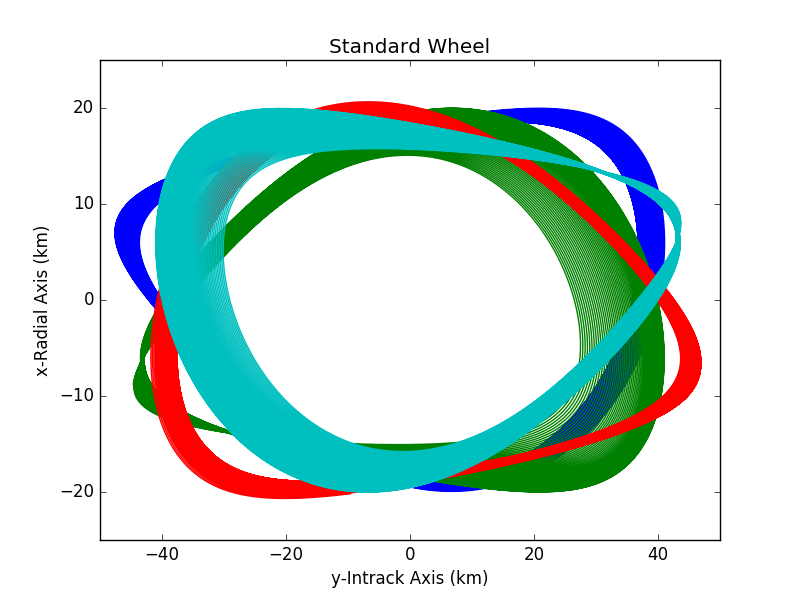}}
	\subfigure{\includegraphics[width=.45\textwidth]{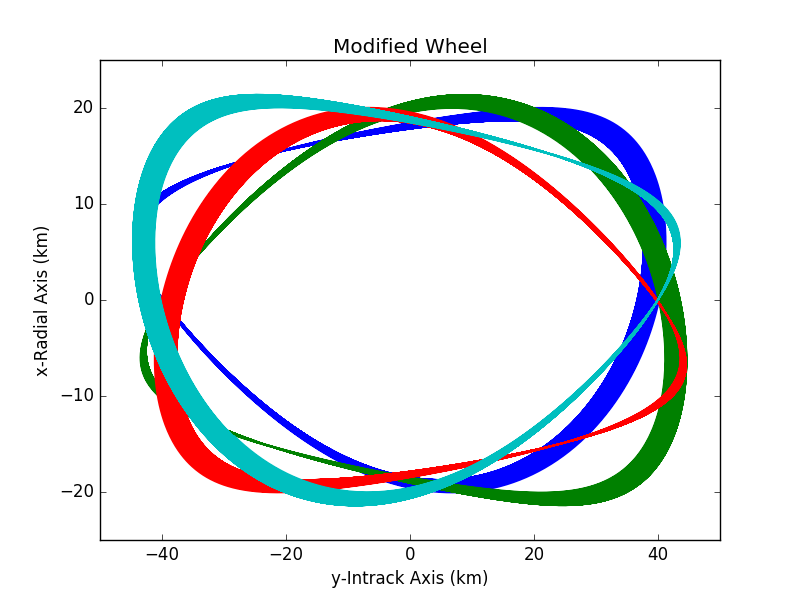}}\\
	\subfigure{\includegraphics[width=.45\textwidth]{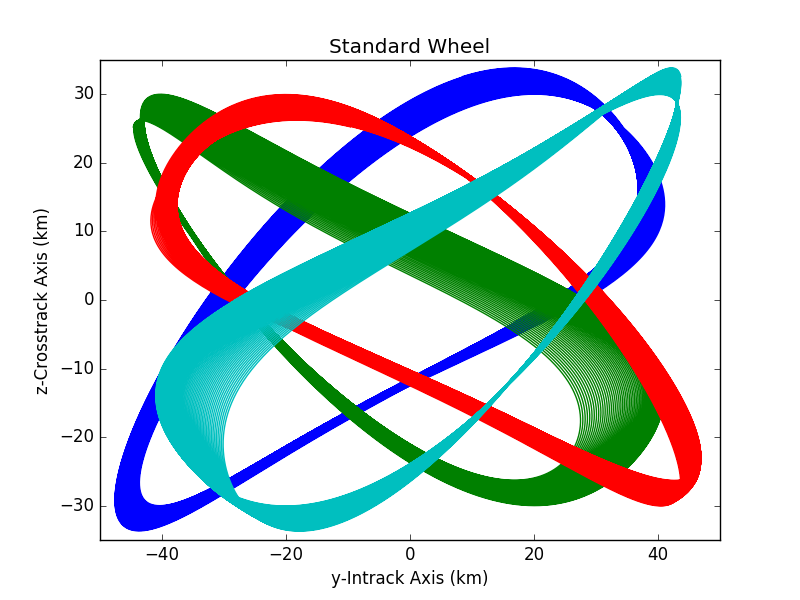}}
	\subfigure{\includegraphics[width=.45\textwidth]{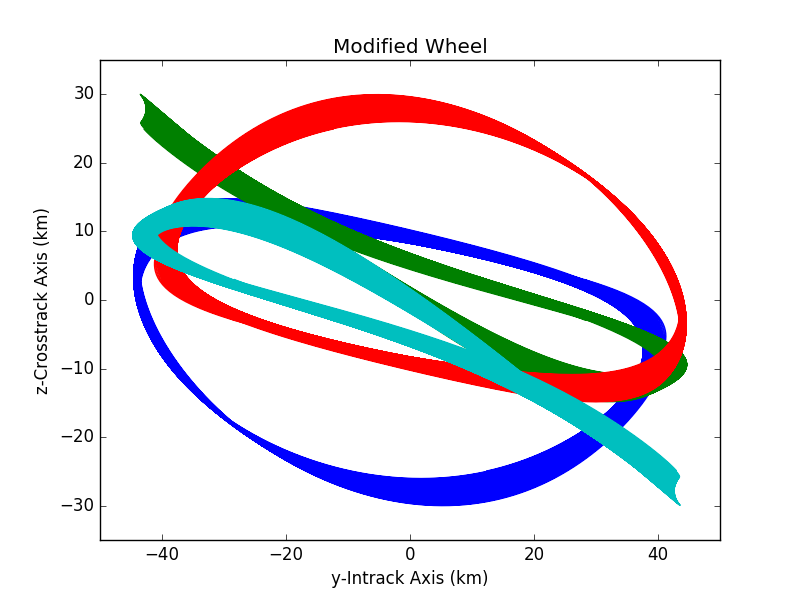}}\\
	\caption{Similar to figure 8, but with propagation for thirty days with a 12 by 12 gravity field as well as solar and lunar gravity.}
\end{figure}

\begin{figure}[hbt!]
	\label{pertmolin2}
	\centering
	\subfigure{\includegraphics[width=.45\textwidth]{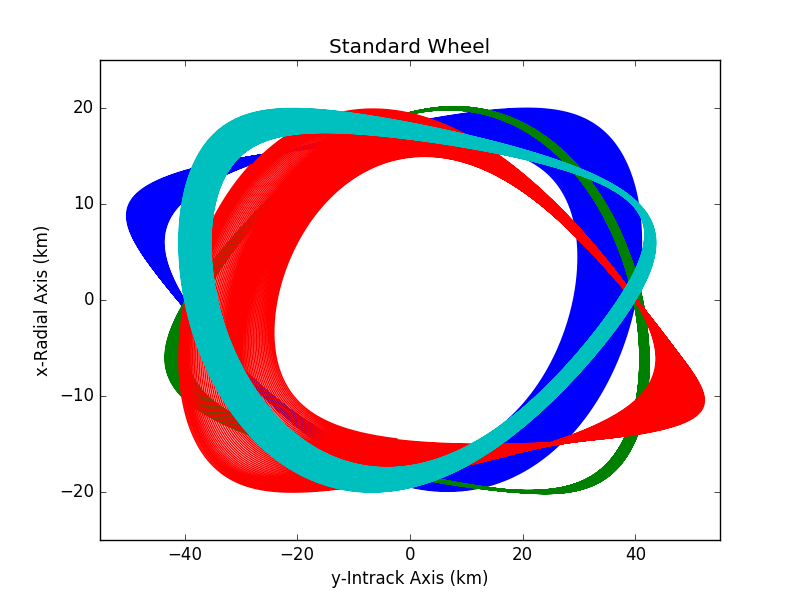}}
	\subfigure{\includegraphics[width=.45\textwidth]{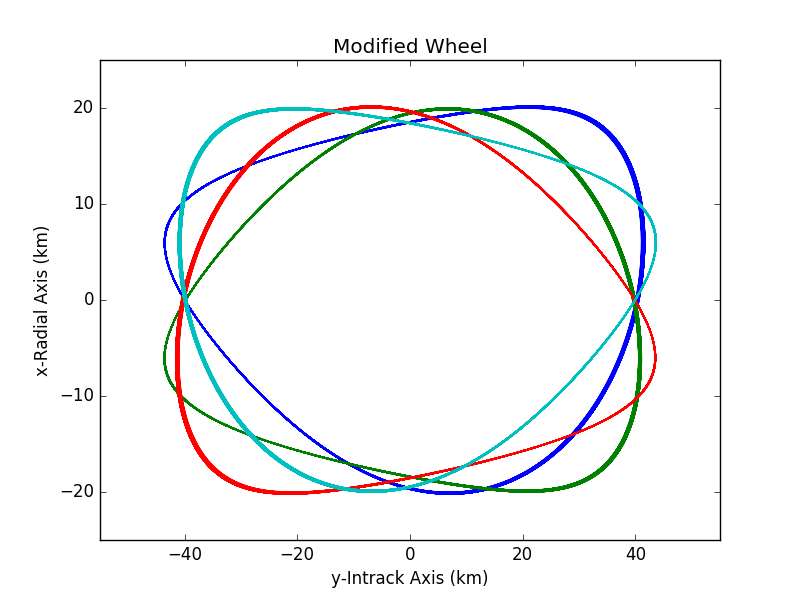}}\\
	\subfigure{\includegraphics[width=.45\textwidth]{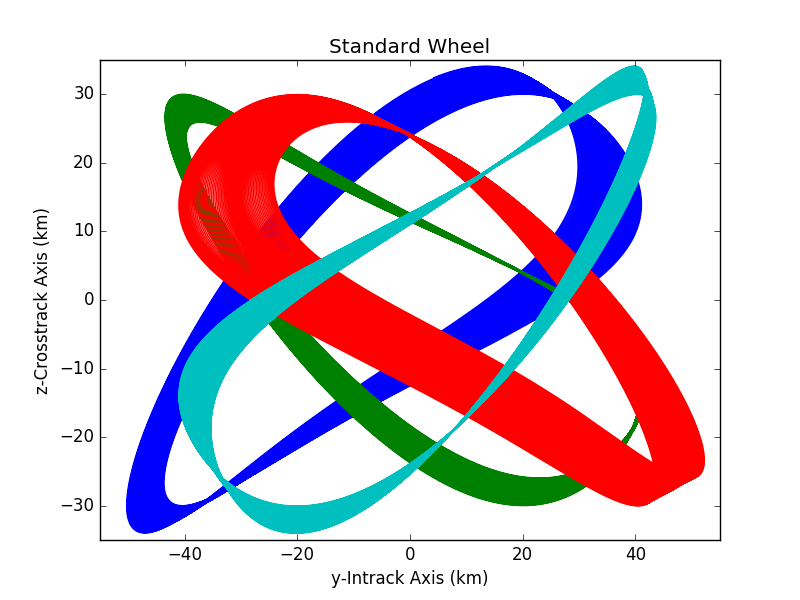}}
	\subfigure{\includegraphics[width=.45\textwidth]{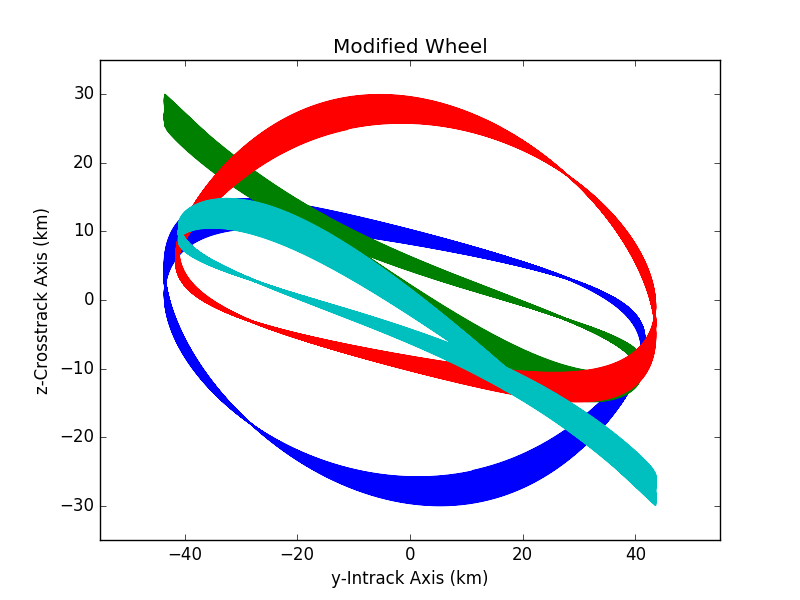}}\\
	\caption{Similar to figure 9, with propagation for thirty days, a 12 by 12 gravity field, solar and lunar gravity, but about a reference orbit with slightly less than a half day period.}
\end{figure}

\section{A Modified Control Algorithm}

Schaub and Alfriend followed their original mean latitude $J_2$-invariance paper with an impulsive control strategy for maintaining a given set of differential mean elements \cite{impulse}. While this control strategy maintains a formation's differential mean elements between chief and deputy regardless of the orbital regime or the formation type, it was specifically optimized for formations designed with the mean latitude constraint in slightly elliptical regimes. The strategy outlined in \cite{impulse} consists of three burns per chief period, two for in-plane control, and a third for cross-track control. To improve upon this strategy for the specific use case presented in this paper, there are two optimizations that can be made.\\

The first optimization is the most general, and leads to improved performance in highly elliptical and inclined orbits. This optimization is in the location of the cross-track burn. Note that this cross-track burn is chosen at one of two true latitudes ($\pi$ radians apart). The given necessary combination of differential inclination and right ascension of the ascending node (along with a corresponding change to the argument of perigee by $-\cos[i]\delta\Omega$) can be adjusted simultaneously at either of these positions

\begin{align}
\theta_c&=\arctan\bigg(\frac{\delta\Omega\sin i}{\delta i}\bigg)\\
\Delta\mathrm{v}_h&=\frac{h}{r}\sqrt{\delta i^2+\delta\Omega^2\sin^2 i} 
\end{align}

Schaub and Alfriend then choose the true latitude angle at which the burn will be in the direction of the positive angular momentum vector. The adoption of this arbitrary convention slightly decreases the complexity of the algorithm, but can significantly increase the delta-v cost of the cross-track burn as compared with the alternative in certain situations. The optimal choice is the true latitude angle at which the chief is farther from the Earth. At this location, a given cross-track burn magnitude will adjust the orbital elements to a greater extent.\\

 Thus, one need only to test the chief distance from Earth at the two true latitudes and then employ the one at which the distance is greater. Afterwards, employ the same exact formula for the burn magnitude. Determining the direction (positive or negative chief angular momentum vector) of the burn then is simple. However, care must be taken in which of two measures are employed to determine it, since depending on the situation, the $\delta i$ or $\delta\Omega$ to be adjusted may be very small.
 
 Given the larger magnitude $\delta i$ or $\delta \Omega$, the sign of the control vector (in the direction of or opposite chief angular momentum) will be
 
 \begin{align}
 &\mathrm{sign} (\delta i\cos\theta_c)\\
 &\mathrm{sign} (\delta\Omega\sin\theta\sin i) \end{align}

This choice of sign is obtained from $u_h$ in equation 1 in \cite{impulse}.

Supposing the true anomaly values for the possible cross-track burn locations are $\nu_c=0,\pi$, then the ratio of the burn magnitude to adjust by the differential elements $\delta i$ and $\delta \Omega$ at the two different locations is $(1+e)/(1-e)$. For $e=0.72$, the ratio is approximately a factor of 6. Note that $(1+e)/(1-e)$ is the maximum this ratio achieves, and in general, the savings from choosing the optimal location are not as high as this ratio.\\

The situation described above is approximately the case for a Molniya orbit with high inclination, 90 or 270 degree argument of perigee, and largely $\delta\Omega$ correction rather than $\delta i$ correction to be achieved as in the case of an in-plane $J_2$-invariant formation. Depending on the reference orbit and the drifts to be adjusted, the original arbitrary choice may coincide with the optimal case, but in others, the cross-track burn could be non-optimal. Given the dependence on reference eccentricity, it is easy to see why the choice is fairly arbitrary for low eccentricity reference orbits.

The second optimization is to the in-plane burns. The original control scheme \cite{impulse} was developed with the idea in mind that the elements to be adjusted, $\delta\omega$ and $\delta M$, would often be nearly equal but opposite. However, this is not the case in the situations arising from the in-plane $J_2$-invariant condition. Instead, due to higher-order gravitational perturbations, the mean anomaly drift often dwarfs the portion of the argument of perigee drift not associated with the right ascension of the ascending node drift (adjusted in the cross-track burn discussed already). As such, it makes sense to revise the control strategy to handle large mean anomaly adjustments more efficiently. This is done by modifying the differential semi-major axis instead, correcting the differential mean anomaly over time.

Consider the problem of employing a single two-burn Hohmann transfer initiated at one of the apses to cause a satellite with mean motion $n$ under a Keplerian force model to move along in its orbit by a mean anomaly of $\delta M$ more than it would have without control in the time it takes to reach the sames apsis again. After the second burn, the satellite will be in some orbit with semi-major axis $a+\delta a$, while all other orbital elements besides mean anomaly will remain unchanged from their starting values. Since the transfer semi-major axis is the arithmetic mean of the initial and final positions, $a_{tp},a_{ta}$, the transfer semi-major axes when the maneuver is begun at perigee, and apogee respectively are

\begin{align}
a_{tp}&=a+\frac{1+e}{2}\delta a\\
a_{ta}&=a+\frac{1-e}{2}\delta a
\end{align}

To maximize the amount of mean anomaly drift achieved in a period, it is best for the change in semi-major axis to be as dramatic as possible from the start so that one iteration of the algorithm should begin with its first burn at perigee. Analysis for the apogee initialized maneuver is similar, but omitted since it is less efficient.\\

Knowing that the semi-major axis is $a_{tp}$ and then $a+\delta a$ for the two parts of the maneuver

\begin{align}
\delta M&=2\pi-n(\tau_1+\tau_2)/2\\
n&=\sqrt{\frac{\mu}{a^3}}\\
\tau_1&=2\pi\sqrt{\frac{a_{tp}^3}{\mu}}\\
\tau_2&=2\pi\sqrt{\frac{(a+\delta a)^3}{\mu}}
\end{align}

After series expansion in terms of $\delta a$, truncation after the linear term, then rearranging for $\delta a$

\begin{equation}
\label{dadm}
\delta a\approx-\frac{4 a\delta M}{3 (3 + e) \pi}
\end{equation}

This will form an additional corrective term to be fed into the original control algorithm for some specified $\delta M$ to be decided.

To reduce the propellant cost, the portion of the mean anomaly drift to be adjusted using a differential semi-major axis drifting approach and the amount directly adjusted by a radial impulse is chosen to reduce the differential argument of perigee adjustment impulses. Let $\delta M$ be the difference between the deputy mean anomaly at the start of one iteration of the algorithm (chief perigee) from the target mean anomaly (the chief's mean anomaly plus the specified differential mean anomaly to maintain). Let $\delta M'$ be the differential mean anomaly to be adjusted directly using a radial burn, then $\delta M-\delta M'$ is the amount of mean anomaly the deputy needs to drift in one period using the method from \ref{dadm}. Choose $\delta M'$ to reduce $|\Delta\mathrm{v}_{rp}|+|\Delta\mathrm{v}_{ra}|$ and $\delta M-\delta M'$:

\begin{align}
\Delta\mathrm{v}_{rp}&=-\frac{na}{4}\bigg[\frac{(1+e)^2(\delta\omega+\delta\Omega\cos i)}{\sqrt{1-e^2}}+\delta M'\bigg]\\
\Delta\mathrm{v}_{ra}&=\frac{na}{4}\bigg[\frac{(1-e)^2(\delta\omega+\delta\Omega\cos i)}{\sqrt{1-e^2}}+\delta M'\bigg]
\end{align}

The first consideration is that if the signs of $\delta M$ and $\delta\omega+\delta\Omega\cos i$ are such that decreasing $|\Delta\mathrm{v}_{rp}|+|\Delta\mathrm{v}_{ra}|$ increases $\delta M-\delta M'$, we select $\delta M'=0$. If both the radial and tangential costs can be reduced by choosing a nonzero $\delta M'$, we consider that $\delta M'\in\big[\frac{(1-e)^2(\delta\omega+\delta\Omega\cos i)}{\sqrt{1-e^2}},\frac{(1+e)^2(\delta\omega+\delta\Omega\cos i)}{\sqrt{1-e^2}}\big]$ results in the same $|\Delta\mathrm{v}_{rp}|+|\Delta\mathrm{v}_{ra}|$, and choose the larger to reduce $\delta M-\delta M'$. However, if this quantity is greater than $\delta M$, we set $\delta M'=\delta M$ to avoid increasing the tangential burn costs. These considerations result in equation \ref{dmprime}.

\begin{equation}
\label{dmprime}
\delta M'=
\begin{cases}
0        & \text{if } \delta M (\delta\omega +\delta\Omega \cos i)> 0 \\
\mathrm{sign}(\delta M)\min\Bigg(|\delta M|,\bigg|\frac{(1+e)^2(\delta\omega+\delta\Omega\cos i)}{\sqrt{1-e^2}}\bigg|\Bigg)        & \text{otherwise}
\end{cases}
\end{equation}


As such, the tangential burns become

\begin{align}
\Delta\mathrm{v}_{\theta p}&=\frac{na\sqrt{1-e^2}}{4}\bigg[\frac{\delta a+\delta a'}{a}+\frac{\delta e}{1+e}\bigg]\\
\Delta\mathrm{v}_{\theta a}&=\frac{na\sqrt{1-e^2}}{4}\bigg[\frac{\delta a+\delta a'}{a}-\frac{\delta e}{1-e}\bigg]\\
\delta a'&=-\frac{4 a(\delta M-\delta M')}{3 (3 + e) \pi}
\end{align}

Where $\delta a$ is the difference between the deputy semi-major axis at the start of one iteration of the algorithm (chief perigee) from the target semi-major axis (the chief's semi-major axis plus the specified differential semi-major axis to maintain).\\

Thus, delta-v can be reduced at the cost of changing the desired differential semi-major axis from original specifications by a small amount, and needing to start the algorithm iterations at perigee.

\begin{figure}[hbt!]
	\label{comp1}
	\setlength{\tempwidth}{.47\linewidth}
	\settoheight{\tempheight}{\includegraphics[width=\tempwidth]{dm_out}}
	\centering
	\hspace{\baselineskip}
	\subfigure{\includegraphics[width=\tempwidth]{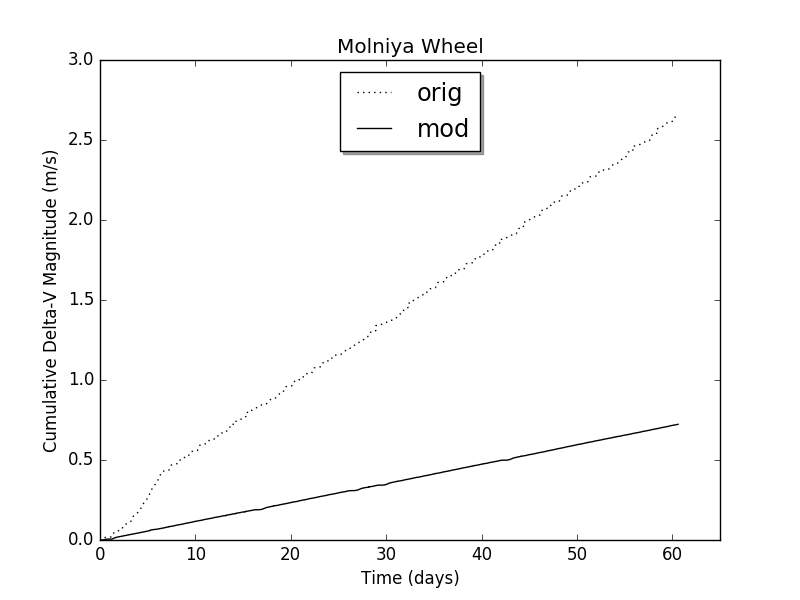}}\hfil
	\subfigure{\includegraphics[width=\tempwidth]{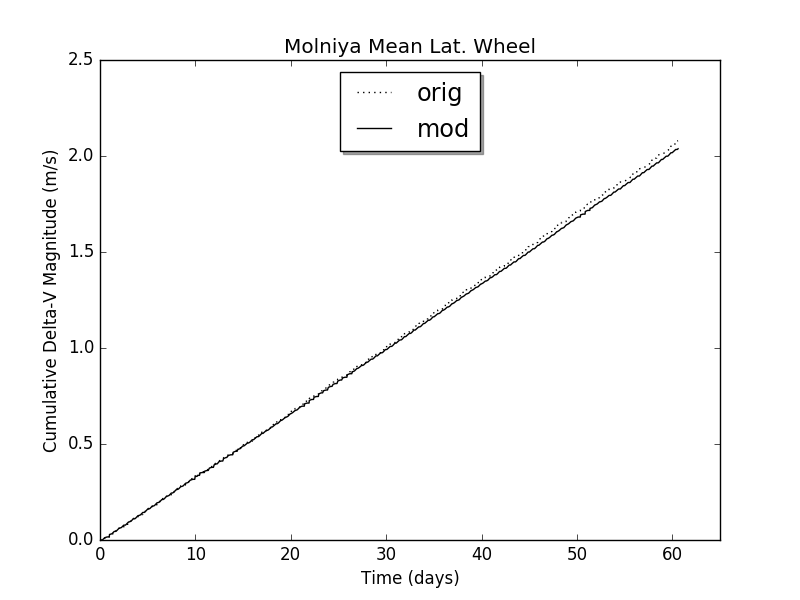}}\\
	\subfigure{\includegraphics[width=\tempwidth]{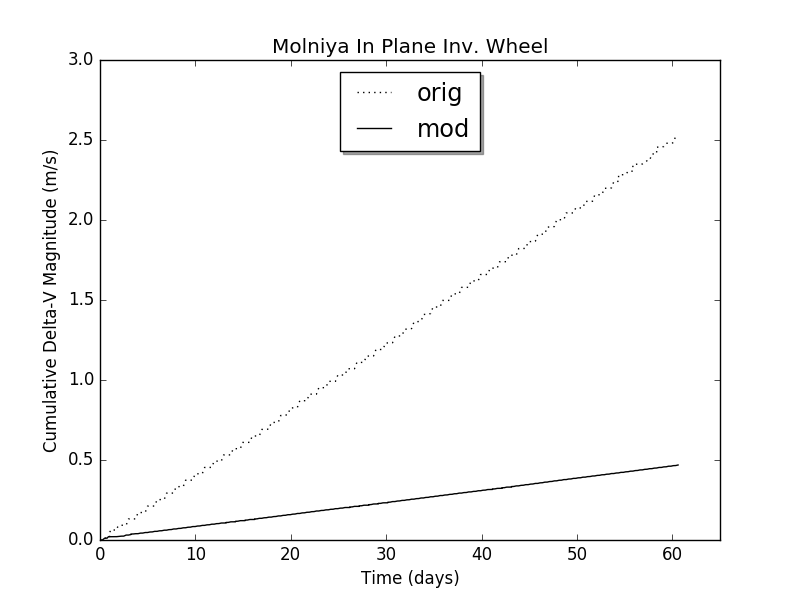}}\hfil
	\subfigure{\includegraphics[width=\tempwidth]{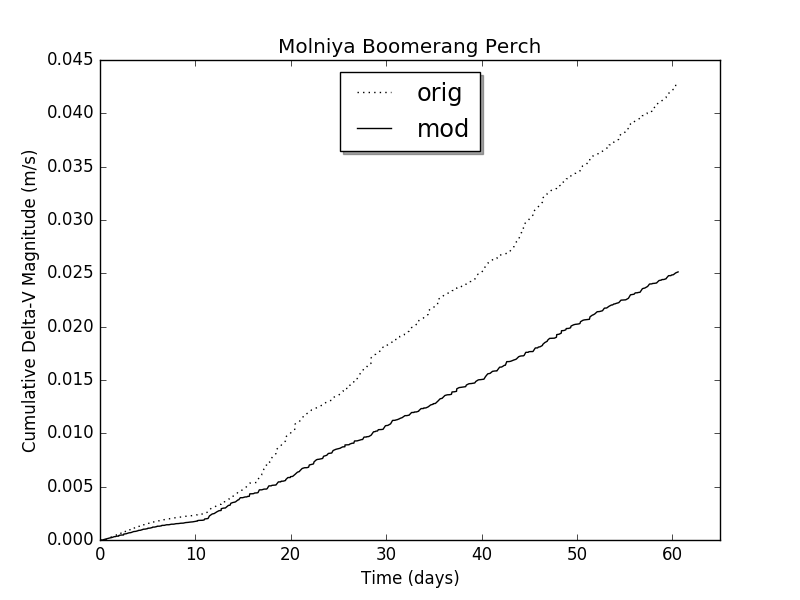}}

	\caption{Plots of the delta-v cost for maintenance by the original and modified control algorithms for various formations over time.}
	\label{fig:figcon}
\end{figure}

\begin{table}[]
	\label{costs}
	\centering
	\begin{tabular}{|l|l|l|l|}
		\hline
		Formation         & Cross-Track Cost (m/s) & Perigee Cost (m/s) & Apogee Cost (m/s) \\ \hline\hline
		Wheel             & 1.952                   & 0.490               & 0.209             \\ \hline
		Modified Strategy        & 0.319                   & 0.367               & 0.0365             \\ \hline\hline
		Mean Lat Wheel    & 0.0569                 & 1.505               & 0.518              \\ \hline
		Modified Strategy         & 0.0151                 & 1.505               & 0.518              \\ \hline\hline
		InPlane Inv Wheel & 2.44                   & 0.0442               & 0.0399              \\ \hline
		Modifed Strategy          & 0.410                   & 0.0420              & 0.0164            \\ \hline\hline
		Boomerang Perch   & 0.00679                & 0.0176              & 0.0184            \\ \hline
		Modified Strategy         & 0.00392                & 0.00982             & 0.0114            \\ \hline
	\end{tabular}
	\caption{A breakdown of the delta-v cost of formation maintenance for the various formation types under the two different maintenance strategies over a two month simulation.}
\end{table}

Examples of delta-v savings are presented in figure \ref{fig:figcon}. In this figure, a chief satellite is initialized at the mean elements $(26561\mathrm{km}, 0.72, 63.4\si{\degree}, 50\si{\degree}, 270\si{\degree}, 180\si{\degree})$, and deputy satellites are placed in wheel, mean latitude wheel, in-plane invariant wheel formations with parameters $D=10\mathrm{km},\beta=-135$ degrees, as well as a non-wheel-type formation called the boomerang perch with $y=10\mathrm{km}$ (see appendix B). These satellites are then propagated for two months under a 12x12 gravity model with solar and lunar gravity. The modified control algorithm is superior in all cases except for the mean latitude wheel where the two algorithms match performance. One expects very little performance change for the mean latitude wheel, since it prevents much $\delta\Omega$ drift, and the $\delta M$ drift is roughly equal and opposite the $\delta \omega$ drift. In this case, the algorithms perform almost exactly the same in-plane burns and the difference in cross-track cost is negligible compared to the in-plane maintenance costs. Table 1 lists the different types of maintenance required, allowing one to identify to what degree performance gains are due to cross-track or in-plane changes to the maintenance strategy.  The in-plane invariant wheel with the modified control algorithm has the lowest cost to maintain of all the wheels. In particular, the cost of in-plane maintenance is an order of magnitude lower for the in-plane $J_2$-invariant formation with the modified algorithm than for any other wheel formation. However, some of these savings are offset by a bump in the out-of-plane maintenance cost.

 Note that while the original formation maintenance algorithm due to \cite{impulse} is designed for slightly elliptical reference orbits, it does suffer issues with singularities at extremely low eccentricities in perturbed environments such as LEO. These same issues are suffered by the algorithm we present here, and stem from maintaining differential mean classical elements. As a chief's mean eccentricity changes over time due to maneuvers or perturbations, a  constant differential mean element set can lead to a different formation geometry or even be unattainable given a negative eccentricity resulting from adding the differential element set to the current chief's element set. Approaches using other differential orbital element sets can remedy this problem \cite{d2010autonomous}. As such, the algorithm discussed here should not be used in slightly eccentric perturbed environments, and will see its best use in moderately to highly elliptical settings where it has significant delta-v savings over existing approaches.

\section{Conclusion}

This paper presents novel contributions to the design and control of satellite formations in highly elliptical orbits, including a comparison with existing methods in different orbital regimes. We developed a condition on differential inclination to yield a satellite formation that exhibits approximately periodic in-plane motion despite the $J_2$ perturbation. Simulations demonstrated the suitability of this condition for constructing wheel formations in highly elliptical orbits. Various means of determining differential orbital elements based on geometric parameters were discussed. An alternative derivation, including a significant correction, was presented for a wheel formation design algorithm by Chao. A summary was given of the highly elliptical wheel algorithm of Sengupta and Vadali, and the two geometric formation design methods were compared. We also offered two modifications to a control algorithm from Schaub and Alfriend to decrease propellant cost to maintain highly elliptical formations. Further, a novel perching formation in the shape of a boomerang exhibiting minor in-track oscillation was presented and analyzed with both control algorithms.

 Future work might include characterizing effects of higher-order gravitational harmonics that exhibit such a significant influence on the relative motion of satellites in half day period resonant Molniya orbits, and attempting to correct these effects in addition to the effects of $J_2$. Bringing more modern dynamical systems theory approaches to bear on highly elliptical formation design could also prove fruitful.

\begin{appendices}
	
	\section{Derivations} \label{a:derivations}
	\subsection{Chao Wheel Algorithm}
	\label{A:wheel}
	For our alternate derivation of the Chao wheel builder algorithm, one must employ a third imaginary satellite (we shall call it the ``virtual chief") in a circular orbit. Since a superposition principle is used to add the relative motions of the virtual chief with respect to the chief and the deputy with respect to the virtual chief, and formulae for the two satellites relative to a circular virtual chief are employed, the two satellites must be fairly low in eccentricity for their circumnavigation formations relative to the virtual chief to be small and their predicted behavior and superposition to be good assumptions. In this derivation, take $e_c, e_d$ to be the eccentricities of chief and deputy respectively, while unsubscripted quantities are assumed with respect to the chief, and differential quantities are between the chief and deputy and equal to the same quantity between virtual chief and deputy.
	
	\begin{equation}
	\vec{r}_{dc}=\vec{r}_{vc}+\vec{r}_{dv}=\vec{r}_{dv}-\vec{r}_{cv}
	\end{equation}
	where subscript $dc$ indicates deputy with respect to chief, $vc$ indicates virtual chief with respect to chief, and so on.
	
	Taking Schaub's equations for a zero eccentricity reference orbit \cite{schaub2004} and approximating what would normally be deputy true anomaly $f$ as deputy mean anomaly $M$, since the deputy and chief, which are viewed as deputies of the virtual chief, are only supposed to be slightly eccentric. All satellites here share the same semi-major axis, and $M$ represents the approximate mean anomaly of both chief and virtual chief:
	
	\begin{align}
	x_{dv}&\approx-a\cos(M+\delta M)e_d\\
	y_{dv}&\approx a\delta(M+\omega)+2a\sin(M+\delta M)e_d\\
	x_{cv}&\approx -a\cos(M)e_c\\
	y_{cv}&\approx 2a\sin(M)e_c
	\end{align}
	
	Supposing for the purpose of centered circumnavigation formations that $\delta M=-\delta\omega$ and subtracting these motions:
	
	\begin{align}
	x_{dc}&\approx-a\cos(M+\delta M)e_d+a\cos(M)e_c\\
	y_{dc}&\approx 2a\sin(M+\delta M)e_d-2a\sin(M)e_c
	\end{align}
	
	Upon expanding all the additive angles inside trigonometric terms:
	
	\begin{align}
	x_{dc}&\approx a\big[\cos M (e_c-e_d\cos\delta M)+\sin M (e_d\sin\delta M)\big]\\
	y_{dc}&\approx 2a\big[\cos M (e_d\sin\delta M) -\sin M (e_c-e_d\cos\delta M)\big]
	\end{align}
	
	Now, recombining terms in a different manner:
	
	\begin{align}
	D&=a\sqrt{\sin^2\delta Me_d^2+(e_c-e_d\cos\delta M)^2}\\
	\sin\beta&=\frac{-\sin\delta Me_d}{D/a}\\
	\cos\beta&=\frac{e_c-e_d\cos\delta M}{D/a}\\
	x_{dc}&\approx D\cos(M-\beta)\\
	y_{dc}&\approx 2D\sin(M-\beta)
	\end{align}
	
	Next, solve for $e_d$ and $\delta M$, given that $\Delta e=\frac{D}{a}$.
	
	\begin{align}
	D&=a\sqrt{\sin^2\delta Me_d^2+(e_c-e_d\cos\delta M)^2}&\implies\\
	D&=a\sqrt{e_d^2+e_c^2-2e_ce_d\cos\delta M}&\implies\\
	\Delta e^2&=e_d^2+e_c^2-2e_ce_d\cos\delta M&\implies\\
	e_d^2&=-e_c^2+\Delta e^2+2e_ce_d\cos\delta M&\implies\\
	e_d^2&=e_c^2+\Delta e^2-2e_c(e_c-e_d\cos\delta M)&\implies\\
	e_d^2&=e_c^2+\Delta e^2-2e_c\Delta e\cos\beta&\implies\\
	e_d&=\sqrt{e_c^2+\Delta e^2-2e_c\Delta e\cos\beta}\\
	\sin\beta&=\frac{-\sin\delta Me_d}{D/a}&\implies\\
	\sin\delta M&=\frac{-\Delta e\sin\beta}{e_d}\\
	\cos\beta&=\frac{e_c-e_d\cos\delta M}{D/a}&\implies\\
	\cos\delta M&=\frac{e_c-\Delta e\cos\beta}{e_d}&\implies\\
	\delta M &= \mathrm{arctan2}\big(-D\sin\beta,ae_c-D\cos\beta\big)
	\end{align}

	\subsection{Higher Eccentricity Formations}

	\subsubsection{Alternative Wheel Formation For High Eccentricities}
	\label{A:highwheel}
	From Sengupta and Vadali \cite{sengupta2007}:
	\begin{align}
	Q_1&=a\sqrt{\delta e^2+\frac{e^2\delta M^2}{1-e^2}}\\
	Q_2&=a(1-e^2)(\delta\omega+\delta\Omega\cos i+\frac{\delta M}{(1-e^2)^{3/2}})\\
	\alpha_0&=\arctan\big(-\frac{\sqrt{1-e^2}\delta e}{e\delta M}\big)
	\end{align}
	To construct a centered wheel formation, set $D=Q_1$ and $Q_2$ such that the in-track axis crossings are equidistant from one another (not equivalent to $Q2=0$) and then solve for the differential orbital elements in terms of $D$ and $\alpha_0$.
	
	To create a formation with radial extrema $D$, and radial phasing (approximate overall phase angle from in-track positive) $\alpha_0$ Sengupta and Vadali provide the following.
	\begin{align} \delta M&=\frac{D}{ae}\cos\alpha_0\sqrt{1-e^2}\\
	\delta e&=-\frac{D\sin\alpha_0}{a}\\
	Q_2&=De\cos\alpha_0
	\end{align}
	
	Given $Q_2$ chosen to center the wheel, solve for $\delta\omega$:
	
	\begin{align}
	\delta\omega+\frac{\delta M}{(1-e^2)^{3/2}}&=\frac{eD\cos\alpha_0}{a(1-e^2)}&\implies\\
	\delta\omega&=\frac{eD\cos\alpha_0}{a(1-e^2)}-\frac{D\cos\alpha_0}{ae(1-e^2)}&\implies\\
	\delta\omega&=-\frac{D\cos\alpha_0}{ae}
	\end{align}

	\subsubsection{Alternative Wheel Formation For High Eccentricities --- Time-centered}
	\label{A:timewheel}
	Begin with the expressions for $\delta e$ and $\delta M$ from the previous section, which are constrained by $D$ and $\alpha_0$. Choose $\delta\omega$ to center the formation in a temporal sense.\\
	
	From Sengupta and Vadali \cite{sengupta2007}, continue with a condition on $Q_2$ that corrects a temporal bias in the formation. Then, equate this with the corresponding expression in terms of orbital elements, substituting in the value of $\delta M$ constrained by the size and phase of the formation, and simplify until $\delta\omega$ is obtained.
	
	\begin{align}
	Q_2&=\frac{De\cos\alpha_0(5-2e^2)}{2+e^2}\\
	Q_2&=a(1-e^2)(\delta\omega+\frac{\delta M}{(1-e^2)^{3/2}})\\
	\delta M&=\frac{D}{ae}\cos\alpha_0\sqrt{1-e^2}&\implies\\
	\delta\omega&=-\frac{2D(1-e^2)}{ae(e^2+2)}\cos\alpha
	\end{align}
	The resulting deputy satellite spends equal times projected onto the positive along-track axis as the negative along-track axis (where along-track is in the direction of chief velocity).
	
	\section{Another Highly Elliptical Formation: The Boomerang Perch}
	
	\begin{figure}[hbt!]
		\centering
		\includegraphics[width=.5\textwidth]{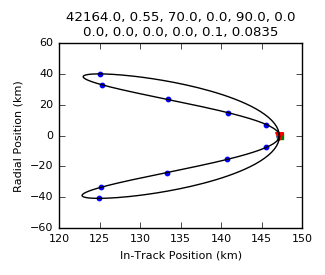}
		\caption{An offset boomerang-shaped orbit resulting from a combination of $\delta\omega$ and $\delta M$. Twelve equally time-spaced blue dots are pictured, while the red triangle represents chief perigee passage, and the green square represents chief apogee passage.}
	\end{figure}
	
	Another formation of interest consists of a combination of $\delta\omega$ and $\delta M$ and attempts to mitigate flight safety problems. This geometry, which we refer to as the ``boomerang perch," chooses the deputy location at chief perigee passage to coincide with its position at chief apogee passage, and as such reduces in-track oscillation for a given distance away from the chief. It is termed a perch for being something of an analogy to an in-track perching (leader-follower) formation in the case of a near circular reference orbit. The boomerang component of the name stems from the resemblance to a boomerang in figure 12.\\
	
	Given the distance $y$ of these two coincident points away from the chief satellite at perigee and apogee, $\delta\omega$ and $\delta M$ can be calculated as follows:
	
	\begin{align}
	\label{boomerang}
	\delta\omega&=\frac{y}{2a}\\
	\delta M&=\sqrt{1-e^2}\delta\omega
	\end{align}
	
	The deputy will sweep across the same radial region twice per chief orbital period: one time quickly around chief perigee, and one time slowly around chief apogee, with the ratio of the two sweep durations higher for higher eccentricity reference orbits. This formation is introduced, in part to demonstrate the efficacy of the modified control algorithm presented on non-wheel-type formations.
	
	The utility of this formation comes from its relatively small ratio of in-track extrema.
	
	\begin{equation}
	\frac{y_f}{y_c}=\frac{1}{\sqrt{1-e^2}}\approx 1+\frac{e^2}{2}
	\end{equation}
	
	Where $y_f$ is the far in-track extrema, and $y_c$ is the close in-track extrema. Compare this to the rectilinear perching behavior induced by $\delta\omega$ alone, or the offset circular behavior induced by $\delta M$ alone, which both have the following ratio.
	
	\begin{equation}
	\frac{y_f}{y_c}=\frac{1+e}{1-e}\approx 1+2e+2e^2
	\end{equation}
	
	From the truncated series expansions, one can see that the boomerang perch's in-track oscillations grow much more slowly than these other two behaviors as a function of increasing reference eccentricity. 
\end{appendices}

\begin{acknowledgements}
The author thanks Dr. George Pollock, Dr. Andrew Rogers, Jamie Wilson, and Matt Schmit for their mentorship, as well as Dr. Josue Munoz for his advice. The author also thanks the manuscript reviewers for their in-depth suggestions to improve the quality of this paper.
\end{acknowledgements}

\bibliographystyle{spmpsci}      
\bibliography{sample}   

\end{document}